\documentclass[aps,prb,reprint,superscriptaddress]{revtex4-2}

\usepackage{footnote}
\usepackage{supertabular,booktabs}
\usepackage{soul}
\usepackage{graphicx}
\graphicspath{{./figures/}}
\usepackage{dcolumn}
\usepackage{bm}

\usepackage{amsmath}
\usepackage{amsfonts}
\usepackage{amssymb}

\usepackage{siunitx}
\usepackage{pifont}
\usepackage{physics}
\usepackage{xparse}

\usepackage{xspace} 

\newcommand\Tstrut{\rule{0pt}{2.4ex}}       
\newcommand\Bstrut{\rule[-1.3ex]{0pt}{0pt}} 
\newcommand{\mathemdash}{\multicolumn{1}{c}{\textemdash}}

\usepackage[colorlinks]{hyperref}

\newcommand{\iso}[1]{\ensuremath{{^{#1}}}}
\newcommand{\tplus}{\ensuremath{^{3+}}\xspace}
\newcommand{\state}[3]{\ensuremath{{}^{#1}\! #2 _{#3}}\xspace}
\newcommand{\yso}{Y$_2$SiO$_5$\xspace}

\newcommand{\eucl}{EuCl$_3$.6H$_2$O\xspace}
\newcommand{\caf}{CaF$_2$\xspace}
\newcommand{\eutrans}{$\eug\rightarrow\eue$\xspace}

\newcommand{\MHz}[1]{\SI{#1}{\mega\hertz}}

\newcommand{\eug}{\state{7}{F}{0}}
\newcommand{\eue}{\state{5}{D}{0}}

\newcommand{\rme}[3]{\ensuremath{\left\langle{#1\vphantom{#2 #3}}\left|\left|{#2}\vphantom{#1 #3}\right|\right|{#3}\vphantom{#1 #2}\right\rangle}}

\newcolumntype{Y}[1]{S[table-format = #1]}

\newcommand{\citeit}[1]{\cite{#1}\xspace} 
\newcommand{\etal}{\textit{et al}.\xspace}

\begin{document}

\title{Complete crystal field calculation of Zeeman-hyperfine splittings in europium}
\author{Kieran M. Smith}
\affiliation{Centre for Quantum Computation and Communication Technology, Research School of Physics, The Australian National University, Canberra, Australia}
\author{Michael F. Reid}
\affiliation{School of Physical and Chemical Sciences, University of Canterbury, PB 4800, Christchurch 8041, New Zealand}
\affiliation{Dodd-Walls Centre for Photonic and Quantum Technologies, New Zealand}
\author{Matthew J. Sellars}
\affiliation{Centre for Quantum Computation and Communication Technology, Research School of Physics, The Australian National University, Canberra, Australia}
\author{Rose L. Ahlefeldt}
\affiliation{Centre for Quantum Computation and Communication Technology, Research School of Physics, The Australian National University, Canberra, Australia}

\date{\today}

\begin{abstract}
Computational crystal-field models have provided consistent models of both electronic and Zeeman-hyperfine structure for several rare earth ions. These techniques have not yet been applied to the Zeeman-hyperfine structure of Eu\tplus because modeling the structure of the $J=0$ singlet levels in Eu\tplus requires inclusion of the commonly omitted lattice electric quadrupole and nuclear Zeeman interactions. Here, we include these terms in a computational model to fit the crystal field levels and the Zeeman-hyperfine structure of the \eug and \eue states in three Eu\tplus sites: the C$_{4v}$ and C$_{3v}$ sites in \caf and the C$_2$ site in \eucl. Close fits are obtained for all three sites which are used to resolve ambiguities in previously published parameters, including quantifying the anomalously large crystal-field-induced state mixing in the C$_{3v}$ site and determining the signs of Zeeman-hyperfine parameters in all three sites. We show that this model allows accurate prediction of properties for Eu\tplus important for quantum information applications of these ions, such as relative transition strengths. The model could be used to improve crystal field calculations for other non-Kramers singlet states. We also present a spin Hamiltonian formalism without the normal assumption of no $J$ mixing, suitable for other rare earth ion energy levels where this effect is important.

\end{abstract}
\maketitle

\section{Introduction}
Crystals containing rare earth ions are now extensively studied for quantum information applications, particularly quantum memories. These applications almost always require precise knowledge of the hyperfine structure of the rare earth ion in the optical ground and excited states employed. In memory protocols such as the gradient echo memory \citeit{Hetet2008a} or the atomic frequency comb protocol \citeit{Afzelius2009}, hyperfine structure information is used to spectrally prepare the memory by pumping population at specific frequencies into non-resonant hyperfine states. Knowledge of the hyperfine structure is also used to predict the locations of ZEFOZ (zero first order Zeeman) \citeit{Fraval2004} points where hyperfine transitions are protected from dephasing caused by magnetic fluctuations, and to identify fields at which $\Lambda$ transitions with favorable oscillator strengths for quantum memories can be obtained \citeit{Guillot-Noel2005, Guillot-Noel2010}.

Hyperfine structure is typically determined from experimental data recorded over a range of magnetic fields, which is then fitted to a spin Hamiltonian that parameterizes any electronic contributions to the nuclear structure. This approach has been used very successfully, for instance, for predicting ZEFOZ points in Eu\tplus:\yso to extend the coherence time by six orders of magnitude \citeit{Zhong2015}. However, there are both limitations and drawbacks to using spin Hamiltonians. They cannot be applied when the electronic contributions to the nuclear structure are not constant, for instance for Kramers rare earth ions over a large range of magnetic fields. The experimental data can also contain many lines, particularly for ions with high nuclear spin or in crystals with multiple magnetic subsites, which can be difficult to fit to a spin Hamiltonian, especially in the case of low symmetry where there are many free parameters. Finally, the spin Hamiltonian often provides little physical insight into the ion or the site.

An alternative is to use a full crystal field model for the whole $4f$ configuration of the rare earth ion, including interactions that generate hyperfine structure. These types of models have been successfully applied for several different hosts doped with Kramers and non-Kramers rare earth ions, including Pr$^{3+}$ \citeit{McLeod1997, Wells1999, Goldner2004, Goldner2008, Guillot-Noel2010}, Er$^{3+}$ \citeit{Popova2000, Gerasimov2016, Horvath2019, Popova2019}, Ho$^{3+}$ \citeit{McLeod1997, Wells2004, Baraldi2007, Mazzera2012, Pytalev2012, Shakurov2014, Boldyrev2019, Mothkuri2021}, Tb$^{3+}$ \citeit{Liu1988, Wells2009}, and Tm$^{3+}$ \citeit{Guillot-Noel2005, Goldner2008}. However, a complete crystal field calculation of Eu\tplus Zeeman-hyperfine structure has not been demonstrated. Terms that are omitted for other ions due to their relatively small contributions become important in the commonly studied $J=0$ crystal field levels in Eu\tplus. These are the electric quadrupole contribution from the lattice field gradient, and the nuclear Zeeman. The qualitative contribution of these terms to the Zeeman-hyperfine structure in Eu\tplus is well-established \citeit{Elliott1957, Sternheimer1966, Blok1966, Silversmith1985f}, but accurate estimates of their exact size in different sites have largely been lacking. Here, we implement a crystal field model with all contributions included, and apply the model to three different Eu\tplus sites: the C$_{4v}$ and C$_{3v}$ sites in \caf, and the C$_2$ site in \eucl. We also extract spin Hamiltonian parameters from the crystal field fit and discuss how crystal field models can be used to resolve ambiguities in the spin Hamiltonian.

\section{Theoretical background}
\subsection{Crystal field Hamiltonian}
Here, we first present the full crystal field Hamiltonian for the $4f$ configuration with expressions for each individual term. Most of these expressions have been covered in detail, \citeit{Wybourne1965, Weiss1978, Liu2005, Guillot-Noel2005, Guillot-Noel2010}, but here we give all terms in a unified notation. 

The complete Hamiltonian appropriate for modeling the $4f^N$ configuration is
\begin{equation}
H = H_\text{FI} + H_\text{CF} + H_\text{Z} + H_\text{HFS}\,. \label{eq:4fn_hamiltonian}
\end{equation}
The terms in Eq. \eqref{eq:4fn_hamiltonian} represent the free-ion, crystal-field, electronic and nuclear Zeeman, and hyperfine interactions respectively.

The free-ion Hamiltonian $ H_\text{FI}$ can be expressed as \citeit{Carnall1989}
\begin{equation}
\begin{aligned}
{ H}_\text{FI} &= E_\text{avg} + \sum_{k=2,4,6} F^k f_k + \zeta_{4f} A_\text{SO} + \alpha L(L+1) \\
& + \beta G(G_2) + \gamma G(R_7) + \sum_{i=2,3,4,6,7,8}T^it_i \\
& + \sum_{i=0,2,4}M^i m_i + \sum_{i=2,4,6}P^i p_i\,. \label{eq:free_ion}
\end{aligned}
\end{equation}
In this equation, $E_\text{avg}$ describes the spherically symmetric part of the Hamiltonian which shifts the entire configuration, $F^k$ and $\zeta_{4f}$ describe the electrostatic and spin-orbit integrals, respectively, and $f_k$ and $A_\text{SO}$ represent the angular parts of the electrostatic and spin-orbit interactions, respectively. The remainder of the terms are smaller. Interactions with other configurations of the same parity as the $4f$ configuration are described by two-body Trees configuration interaction parameters $\alpha$, $\beta$ and $\gamma$, where $L$ is the total angular momentum, and $G(G_2)$ and $G(R_7)$ are the Casimir operators \citeit{Rajnak1963} for the groups $G_2$ and $R_7$ \citeit{Judd1998}. The three-body interaction parameters $T^i$ and three-body operators $t_i$ are included for systems of more than two electrons and account for couplings of the $f^N$ states to states in higher energy configurations via Coulomb interactions \citeit{Judd1966}. Finally, two relativistic corrections are included, the Marvin integrals $M^i$ for spin-spin and spin-other-orbit effects \citeit{Marvin1947}, and the two-body electrostatically correlated magnetic configuration interaction parameters $P^i$ \citeit{Judd1968}, with $m_i$ and $p_i$ being the associated operators. Methods for calculating matrix elements using reduced matrix elements tabulated by Nielson and Koster \citeit{Nielson1963} have been covered in detail \citeit{Weiss1978}. Mean free-ion parameters have been tabulated for the rare-earth ions \citeit{Gorller-Walrand1996a}, however it is necessary to allow $E_\text{avg}$, the Slater parameters $F^k$, and spin orbit parameter $\zeta_{4f}$ to vary during fitting to experimental levels as these parameters have the most significant effect on the calculated spectra. 

The crystal field Hamiltonian $H_\text{CF}$ can be written as
\begin{equation}
{ H}_\text{CF} = \sum_{k = 2,4,6}\sum^{k}_{q=-k}{{B}^k_q {C}^{(k)}_{q}}\,,
\end{equation}
where ${B}^k_q$ are the crystal field expansion coefficients, and $C^{(k)}_q$ are spherical tensor operators using Wybourne's normalization of the spherical harmonics ${Y}_{kq}$ \citeit{Wybourne1965}:
\begin{equation}
{C}^{(k)}_q = \sqrt{\frac{4\pi}{2k+1}}{Y}_{kq}\,.
\label{eq:sphtensor}
\end{equation}
The $B^{k}_q$ with non-zero $q$ may be complex in low symmetries, and in that case the real and imaginary parts must be considered as independent, real parameters. The number of parameters depends on the point symmetry of the rare-earth ion site. Table IV in the Supplementary Materials lists the non-zero parameters for each of the crystallographic point groups. We note that there are a variety of conventions in the literature for labeling the real and imaginary parts \citeit{Newman1989}.

In the presence of an external magnetic field $\mathbf{B}$, the Zeeman Hamiltonian $H_\text{Z}$ is \citeit{Elliott1957}
\begin{align}
{ H}_\text{Z} ={}&  H_{eZ} + H_{nZ}\,,\\ \intertext{with}
H_{eZ}        ={}&  \mu_B \mathbf{B}\cdot\left(\mathbf{L}+g_s\mathbf{S}\right)\,, \\
H_{nZ}        ={}& -g_n\mu_N\mathbf{B}\cdot\mathbf{I}\,,
\end{align}
where $H_{eZ}$ is the electronic Zeeman interaction and $H_{nZ}$ the nuclear Zeeman interaction. $\mathbf{L}$ and $\mathbf{S}$ are the total electronic orbital and spin angular momentum operators, $g_s$ is the electron spin $g$-factor, $\mu_B$ is the Bohr magneton, $\mu_N$ is the nuclear magneton, and $g_n$ is the nuclear $g$-factor. Matrix elements for these terms are given in Refs. \citeit{Guillot-Noel2005, Guillot-Noel2010}. 

The calculation of hyperfine structure is treated by a multipole expansion of the interactions between the electronic and nuclear states. Generally, only interactions involving the magnetic dipole moment and electric quadrupole moment are considered, as higher order multipoles have a negligible contribution in comparison \citeit{Wybourne1965}. The hyperfine Hamiltonian $H_\text{HFS}$ is then expressed as
\begin{equation}
H_\text{HFS} = H_\text{MD} + H_\text{Q}\,,
\end{equation}
where $ H_\text{MD}$ and $ H_\text{Q}$ are the contributions from magnetic dipole and electric quadrupole moment interactions respectively.

The magnetic dipole contribution is given by \citeit{Judd1998}
\begin{align}
{H}_\text{MD} &= a_l \sum_{i}\mathbf{N}_i\cdot\mathbf{I}\,,\\
\mathbf{N}_i &= \mathbf{l}_i - \sqrt{10}(\mathbf{s}\mathbf{C}^{(2)})_i^{(1)}\,,
\end{align}
where $\mathbf{l_i}$ and $\mathbf{s_i}$ are the orbital and spin angular momentum of the electron $i$, and $\mathbf{I}$ is the nuclear spin operator. The magnetic dipole hyperfine parameter $a_l$ is given by \citeit{Wybourne1965}
\begin{equation}
a_l = 2\mu_Bg_n\mu_N \frac{\mu_0}{4\pi}(1-R)\left\langle{r_e^{-3}}\right\rangle\,,
\end{equation}
where $\mu_B$ is the Bohr magneton, $\mu_0$ is the vacuum permeability, $R$ is a shielding factor describing the effect of the induced closed shell quadrupole moment on the $4f$ electrons \citeit{Sternheimer1966, Sternheimer1967, Sternheimer1986}, and $\expval{r_e^{-3}}$ is the average inverse-cube radius of the $4f$ orbital. Whilst $a_l$ can be related to physical constants, during crystal field fitting it is treated as a free parameter due to the shielding factor $R$.

The electric quadrupole contribution arises from the interaction between the nuclear quadrupole moment and field gradient at the nucleus, which has three sources: the lattice, the $4f$ electrons, and the spherically symmetric closed shells \citeit{Weiss1978}. The latter is included as shielding and antishielding coefficients in the former two terms. Thus, the quadrupole can be written:
\begin{equation}
{H}_\text{Q} = { H}_\text{Q}^\text{lat} + { H}_\text{Q}^{4f}\,,
\end{equation}
where ${ H}_\text{Q}^\text{lat}$ is the lattice contribution and ${H}_\text{Q}^{4f}$ is the $4f$ contribution.
The lattice contribution is given by
\begin{equation}
{H}_\text{Q}^\text{lat} = -\sum_{q=-2}^{2}(-1)^q\left(\frac{\left(1-\gamma_\infty\right)}{\left(1-\sigma_2\right)\left\langle r^2_e\right\rangle}B^2_q\left({r}^2_n\left({C}_n\right)^{(2)}_q\right)\right)\,,\label{eq:lat_quad}
\end{equation}
where $(C_n)_q^{(2)}$ is a spherical tensor operator (Eq. \eqref{eq:sphtensor}) operating on the nuclear wave function, $r_n$ is the nuclear radius, $\expval{r_e^2}$ and is the mean square radius of the $4f$ orbital. $\gamma_\infty$ is the quadrupole anti-shielding parameter accounting for the effect of the quadrupole moment induced on the closed shells by the nucleus, while $\sigma_2$ describes the shielding of the lattice field by the filled $5s$ and $5p$ shells. The $4f$ contribution is given by
\begin{equation}
\begin{aligned}
{H}_\text{Q}^{4f} ={}& \frac{-e^2}{4\pi\epsilon_0}\sum^2_{q=-2}(-1)^q
                       \left({r}^2_n\left({C}_n\right)^{(2)}_q\right)\\
                     & \times\left(1-R\right)\left\langle r^{-3}_e\right\rangle
                       \sum_i\left({{C}_{ei}}\right)^{(2)}_{-q}\,, \label{eq:4fquad}
\end{aligned}
\end{equation}
where $(C_{ei})_{-q}^{(2)}$ is a spherical tensor operator (Eq. \eqref{eq:sphtensor}) operating on the wave function of the electron $i$, and $e$ and $\epsilon_0$ are the electron charge and vacuum permittivity respectively. Matrix elements for the lattice and $4f$ quadrupole terms are provided in Refs. \citeit{Guillot-Noel2005, Guillot-Noel2010}.

In our implementation of the crystal field model, the quadrupole interactions are parameterized using rank two unit tensor operators $\mathbf{U}$ as 
\begin{align}
{H}_\text{Q}^{lat} ={}& \sum_{q=-2}^{2} N^2_q\left(U_n\right)^{(2)}_q\,,\\
{H}_\text{Q}^{4f}  ={}& E_Q\frac{1}{2} \left(\frac{(I+1)(2I+1)(2I+3)}{I(2I-1)}\right)^{\frac{1}{2}}\mathbf{U}^{(2)}_n\cdot\mathbf{U}^{(2)}_e\,,
\end{align}
where $\mathbf{U}^{(2)}_e$ operates on the electronic part of the wave function and $\mathbf{U}^{(2)}_n$ operates on the nuclear part. The strength of the interactions is scaled by the fitted parameters $N^2_q$ for the lattice contribution and $E_Q$ for the $4f$ contribution. An approximate relationship between the $N^2_q$ and $B^2_q$ parameters may be derived by assuming that both may be estimated from the point-charge lattice potential, corrected by relevant shielding factors: $(1-\gamma_\infty)$ for the nucleus, and $(1-\sigma_2)$ for the $4f$ electrons. By including the relevant reduced matrix elements we may write:
\begin{align}
N^2_q ={}& \frac{-(1-\gamma_{\infty})}{(1-\sigma_2)\left\langle r^2_e\right\rangle} \frac{Q}{2}
           \left(\frac{(I+1)(2I+1)(2I+3)}{I(2I-1)}\right)^{\frac{1}{2}} B^2_q\,,\label{eq:lat_cf_param}\\
E_Q   ={}& \frac{-e^2 Q}{4 \pi \epsilon_0}(1-R)\left\langle r_e^{-3}\right\rangle\,,
\end{align}
Here the nuclear quadrupole moment Q is introduced via
\begin{equation}
\rme{I}{r^2_n C^{(2)}_n}{I} = \frac{Q}{2} \left(\frac{(I+1)(2I+1)(2I+3)}{I(2I-1)}\right)^{\frac{1}{2}}\,.
\end{equation}
However, since the shielding/antishielding factors $R, \gamma_\infty$ and $\sigma_2$ do vary between materials, $N_q^2$ and $E_Q$ were free fitting parameters in the model bounded by the known ranges of the shielding/antishielding factors \citeit{Radlinski1986}. Additionally, due to the dependence of $N_q^2$ on the crystal field, this parameter was also bounded by the ratio and signs of the $B^2_q$ parameters.

In summary, the full crystal field Hamiltonian in Eq. \eqref{eq:4fn_hamiltonian} for the $4f$ configuration can be expanded as
\begin{equation}
H = H_{FI}+H_{CF}+(H_{eZ}+H_{nZ})+(H_{MD}+H_Q^{lat}+H_Q^{4f})\,,
\label{eq:longhamil}
\end{equation}
with expressions for each term given above, along with references to tabulations of the corresponding matrix elements in $LS$ coupling. In principle, numerically fitting this equation to a particular rare earth ion requires $49$ total parameters for C$_1$ symmetry. In practice however, the number of parameters allowed to vary in a fit is generally far smaller, as many of parameters are restricted to the average values or ratios calculated from a number of host crystals.

\subsection{Previous modeling of Zeeman-hyperfine structure}
Although most previous crystal field modeling for rare earth crystals has considered only $H_{FI}$ and $H_{CF}$, several workers have included certain of the hyperfine and Zeeman terms in Eq. \eqref{eq:longhamil}. However, these calculations have typically omitted the lattice nuclear quadrupole and nuclear Zeeman interactions based on the assumption that these are negligible contributions compared to the magnetic dipole, $4f$ quadrupole, and electronic Zeeman effects. To understand when this assumption may be valid, it is useful to examine relative sizes of the various contributions in different types of electronic states.


For the Kramers doublets in ions such as Er\tplus, the hyperfine splittings are dominated by the electron-nuclear magnetic dipole interaction with $\sim90\%$ of the observed structure being attributed to
this interaction \citeit{Gerasimov2016, Chen2018}, and the majority of the remainder from the electron-nuclear $4f$ quadrupole interaction. This is due to the true doublet states of Kramers ions which directly allow a large first order magnetic dipole contribution. This generally holds true for any of the Kramers ions \citeit{Macfarlane1987}. Therefore, accurate fits to crystal field levels and hyperfine structure can be achieved by incorporating only the magnetic dipole and $4f$ quadrupole contributions, as has been demonstrated for Er\tplus in several materials including LiYF$_4$ \citeit{Popova2000, Gerasimov2016}, YPO$_4$ \citeit{Popova2019}, and both sites of \yso \citeit{Horvath2019}.

Similar to the Kramers doublets, the non-Kramers doublets and pseudo-doublets are also largely dependent on the magnetic dipole interaction. In true doublets such as the high symmetry (C$_{4v}$) Pr\tplus:\caf ground state \citeit{Wells1999}, there is again a large first order magnetic dipole term. In pseudo-doublet ground states (two singlets split by only a few \si{\per\centi\metre}) such as those observed in Ho\tplus \citeit{McLeod1997, Wells2004, Mothkuri2021} or Tb\tplus \citeit{Liu1988, Wells2009}, it is the second order magnetic dipole interaction, commonly referred to as the pseudo-quadrupole (see Section \ref{sec:spinhamil}), that has the largest contribution. Provided the crystal field splitting of the pseudo-doublet ground state is accurately modeled, satisfactory fits to the hyperfine splittings of transitions can be achieved by calculating the magnetic dipole interaction, with the $4f$ quadrupole having only a minor effect on the splittings \citeit{Wells2004}. Sufficiently accurate splittings of these pseudo-doublets can be achieved by the inclusion of the two-electron correlation crystal field parameters \citeit{McLeod1997}, or by manual manipulation of the levels \citeit{Mothkuri2021}. Hyperfine structure of non-Kramers doublets and pseudo-doublets has been successfully calculated in several hosts, including \caf \citeit{McLeod1997, Wells1999, Wells2004, Wells2009}, SrF$_2$ \citeit{Wells1999, Wells2009}, LiYF$_4$ \citeit{Liu1988, Boldyrev2019}, KY$_3$F$_{10}$ \citeit{Pytalev2012}, CaWO$_4$ \citeit{Shakurov2014}, YAl$_3$(BO$_3$)$_4$ \citeit{Baraldi2007}, YPO$_4$ \citeit{Mazzera2012}, and both sites of \yso \citeit{Mothkuri2021}.

For non-Kramers singlet states in $J\neq 0$ multiplets, the largest contribution to the hyperfine splittings can vary depending on the magnitude of the splittings to nearby crystal field levels within the multiplet. If this splitting is $>\SI{100}{\per\centi\metre}$, the level mixing is typically too small to allow a significant pseudo-quadrupole contribution to the hyperfine structure. In these cases, the largest contribution is the $4f$ quadrupole, but the lattice quadrupole may also need to be considered. For example, hyperfine structure was calculated the $\state{1}{D}{2}(0)$ state of Pr\tplus in LiYF$_4$ \citeit{Goldner2004, Goldner2008} by including $4f$ quadrupole and pseudo-quadrupole terms, but a 13\% discrepancy with experimental data was observed. This is likely due to a significant lattice quadrupole contribution, which was shown to be important in Pr\tplus:La$_2$(WO$_4$)$_3$ \citeit{Guillot-Noel2010, Lovric2011a}, where the lattice quadrupole contributed $\sim25\%$ of the total hyperfine splitting of $\state{1}{D}{2}(0)$.


For non-Kramers singlets with $J=0$, such as \eug and \eue in Eu\tplus, the situation is quite different. $H_Q^{lat}$ and $H_{nZ}$ are the dominant sources of the hyperfine splittings as $H_{eZ}$, $H_{MD}$ and $H_Q^{4f}$ have zero matrix elements for $J=0$ states. It is only through $J$-mixing induced by the crystal field that these terms contribute at all. In these cases, an accurate crystal field calculation is important as the magnitudes of the contributions are sensitive to the splittings to nearby multiplets. In summary, the lattice quadrupole contribution is most significant in non-Kramers singlet states, especially those of $J=0$, and can be ignored in Kramers ions and non-Kramers doublets and pseudo-doublets.


We now turn to the second of the small hyperfine terms, the true nuclear Zeeman term. This is nearly always considered unimportant as the enhancement of the nuclear moment due to the magnetic dipole (the pseudo-nuclear effect, see Section \ref{sec:spinhamil}) is several orders of magnitude larger than the bare nuclear moment in most situations \citeit{Bleaney1973}. However, the true nuclear Zeeman effect becomes important in $J=0$ multiplets, where the pseudo-nuclear term is small, or in cases where the crystal field mixing results in a large pseudo-nuclear Zeeman component in only one direction, such as Tm$^{3+}$:Y$_3$Al$_5$O$_{12}$ \citeit{Guillot-Noel2005}.

We finish by highlighting an important point for quantum information applications, where the crystal field models are often used to predict observables dependent on the wave functions output from the model (such as relative transition probabilities), not just the energy levels. To accurately reproduce these observables, Guillot-No\"el \etal showed that it is important to use the true site symmetry \citeit{Guillot-Noel2010}, rather than to take the common approach of refining the crystal field parameters of low symmetry sites in slightly higher symmetry (such as C$_{2v}$ for C$_2$ or C$_1$). This approach has been widely, and successfully, applied for predictions of crystal field energy level structure, either to reduce the computational complexity or because a lack of data means the lower symmetry crystal field fit is underdetermined. However, it is precisely the mixing caused by the lower symmetry that becomes important for transition probabilities (for instance), so the full symmetry must be used.


Here, we use the full crystal field model and the true site symmetry of three different Eu$^{3+}$ sites to determine quantitative contributions of different terms to the hyperfine splitting and relative oscillator strengths. While the qualitative contributions to Eu$^{3+}$ hyperfine splittings are well known \citeit{Elliott1957, Judd1962, Blok1966, Silversmith1985f}, the lack of a full crystal field model has prevented quantitative determination of the relative size of different contributions. This makes Eu\tplus an excellent test-bed for the full crystal field model. 

\section{Spin Hamiltonian theory}
As described above, the first goal of this paper is to provide accurate crystal field models for Eu\tplus sites. The second goal, addressed in this section, is to provide a direct relationship between the crystal field model and the phenomenological spin Hamiltonian models commonly used in experimental studies of hyperfine structure of individual crystal field levels. The widely used existing expression \citeit{Teplov1967} suitable for other non-Kramers singlet states is inappropriate for the $J=0$ levels of Eu\tplus because it ignores $J$ mixing effects. A correct, crystal field derived expression both allows a physical interpretation of the spin Hamiltonian parameters and resolves certain parameter ambiguities that arise when fitting phenomenological models.

\subsection{Phenomenological spin Hamiltonian}
We first consider the experimental approach to the spin Hamiltonian, and highlight the ambiguities that exist when fitting this Hamiltonian to data. For an electronic singlet state, the appropriate phenomenological spin Hamiltonian contains operators for the nuclear spin only:
\begin{equation}
H_{eff} = \mathbf{(B\cdot Z\cdot B)1}+\mathbf{B\cdot M\cdot I}+\mathbf{I\cdot Q\cdot I}\,,
\end{equation}
where $\mathbf{B}$ is the magnetic field, $\mathbf{I}$ is the nuclear spin operator, $\mathbf{1}$ is the identity operator, $\mathbf{Z}$ is the quadratic Zeeman tensor, $\mathbf{M}$ the linear Zeeman tensor, and $\mathbf{Q}$ the traceless enhanced quadrupole tensor. These three symmetric tensors are unique to that electronic state, and are only constrained by the site symmetry. The most general form of these tensors (applicable to C$_1$ symmetry) is:
\begin{align}
\mathbf{Z} ={}& \mathbf{R}_Z\cdot\begin{bmatrix}
                Z_x & 0   & 0\\
                0   & Z_y & 0\\ 
                0   & 0   & Z_z
                \end{bmatrix}\cdot\mathbf{R}_Z^T\label{eq:sh_Z}\,,\\
\mathbf{M} ={}& \mathbf{R}_M\cdot\begin{bmatrix}
                g_x & 0   & 0\\
                0   & g_y & 0\\
                0   & 0   & g_z
                \end{bmatrix}\cdot\mathbf{R}_M^T\label{eq:sh_M}\,,\\
\mathbf{Q} ={}& \mathbf{R}_Q\cdot\begin{bmatrix}
                -E-\frac{1}{3}D & 0 & 0\\
                 0 & E-\frac{1}{3}D & 0\\
                 0 & 0 & \frac{2}{3}D
                \end{bmatrix} \cdot\mathbf{R}_Q^T\,, \label{eq:sh_Q}
\end{align}
where $\mathbf{R}_i = R(\varphi_i,\theta_i,\psi_i)$ is an Euler rotation matrix. In C$_2$ symmetry, the lowest symmetry studied here, $\varphi = \theta = 0$ for all tensors when the laboratory frame $z$ is aligned with the C$_2$ axis. For axial symmetries such as C$_{4v}$ and C$_{3v}$, all tensors are aligned, $g_x=g_y$, and $E = 0$. 

Experimentally, the spin Hamiltonian parameters required for the particular site symmetry are fit to hyperfine splitting data for a variety of magnetic field values, typically obtained using nuclear magnetic resonance, optically detected nuclear magnetic resonance, or Raman heterodyne spectroscopy. Such a fit furnishes a set of non-unique spin Hamiltonian parameters, because the spin Hamiltonian is insensitive to certain transformations. For instance, in C$_2$ symmetry the Hamiltonian is insensitive to the sign of $\mathbf{Z}$, $\mathbf{M}$, $\mathbf{Q}$, $g_z$, and $D$, and the sign of $E$ is ill-defined since it can be reversed by rotating the entire spin Hamiltonian \ang{90} about $z$. 

Further, there are three possible sets of parameters corresponding to different, equivalent choices of the quantization axis of the quadrupole term. Any one set accurately fits the spin Hamiltonian of a single electronic state, but problems arise when trying to calculate parameters that depend on spin Hamiltonians of two different electronic states, such as relative oscillator strengths. The relative optical transition probability $P_{ij}$ of transitions between different hyperfine states in ground ($i$) and excited ($j$) levels can be calculated from the overlap of the nuclear wave functions $\Psi^n$ determined from the crystal field or spin Hamiltonians of the two levels:
\begin{equation}
P_{ij} = \abs{\braket{\Psi^n_{gi}}{\Psi^n_{ej}}}^2\,.\label{eq:oscillator_strengths}
\end{equation}
However, this is only true if the same quadrupole quantization axes have been used for both spin Hamiltonians.

Finally, a further difficulty arises in materials with multiple magnetically inequivalent subsites, which occur in materials where the rare earth site symmetry is lower than the crystal symmetry. While the spin Hamiltonian of all subsites can be generated from the spin Hamiltonian for one site using the symmetry operations of the crystal, the same base subsite must be correctly chosen for every electronic level. When there is no obvious relationship between spin Hamiltonian parameters in different electronic states, identifying the same subsite can be very difficult. This was seen in Pr\tplus:YAlO$_3$ \citeit{Lovric2012a}, which has two magnetic subsites related by a C$_2$ rotation: the fitted spin Hamiltonian parameters for the \eug and \eue levels in \citeit{Lovric2012a} are for different subsites. 

Additional measurements, such as the relative optical oscillator strength measurements described above, can resolve some but not all of these ambiguities. A crystal field-based spin Hamiltonian, in contrast, can resolve all physically significant spin Hamiltonian parameter ambiguities.

\subsection{Crystal-field-based spin Hamiltonian}\label{sec:spinhamil}
A spin Hamiltonian for a particular electronic state may be extracted from the full crystal-field Hamiltonian
by projecting the crystal-field Hamiltonian into the spin Hamiltonian basis. This approach was used in Ref. \citeit{Horvath2019} and Ref. \citeit{Jobbitt2021}. For a general discussion of projecting Hamiltonians into a smaller basis, see Ref. \citeit{Reid2009}. The projection approach allows the direct calculation of the $\mathbf{Z}$, $\mathbf{M}$, and $\mathbf{Q}$ parameters, by comparing the projected Hamiltonian with the  spin Hamiltonian matrix. Crucially, this method allows a unique set of spin Hamiltonian parameters to be determined with all sign and rotational ambiguities removed.

It is also possible to use a perturbation-theory approach to approximate the relationship between a crystal-field calculation and the spin Hamiltonian. This approach is very helpful for understanding the sizes of the various contributions. The spin Hamiltonian for a non-Kramers singlet state, ignoring $J$-mixing, is commonly written as \citeit{Teplov1967, Macfarlane1987}:
\begin{align}
{ H} ={}& -\mathbf{B}\cdot\left(g_J^2\mu_B^2\mathbf{\Lambda}\right)\cdot\mathbf{B}
        -\mathbf{B}\cdot\left(g_n\mu_N\mathbf{1}
          -2A_Jg_J\mu_B\mathbf{\Lambda}\right)\cdot \mathbf{I}\nonumber\\
        & -\mathbf{I}\cdot\left(A_J^2\mathbf{\Lambda}+\mathbf{T}_Q\right)\cdot\mathbf{I}\,, \label{eq:non-Karmers_SH}
\end{align}
where $g_J$ is the Land\'{e} $g$ value, and $A_J$ is the hyperfine interaction parameter for that state. The lattice and $4f$ quadrupole contributions are combined into a single true quadrupole term $\mathbf{T_Q}$ and $A_J^2\mathbf{\Lambda}$ accounts for the pseudo-quadrupole contribution arising from the magnetic dipole interaction. The contributions from other electronic states are encapsulated by the tensor $\mathbf{\Lambda}$ which sums the interaction with all other states in the multiplet:
\begin{equation}
\Lambda_{ij} = \sum_{n=1}^{2J+1}\frac{\left\langle 0\left|J_i\right|n\right\rangle\left\langle n\left|J_j\right|0\right\rangle}{E_n-E_0}\,,
\end{equation}
 with $\ket{0}$ the singlet state of interest. 

When dealing with only the hyperfine splittings, the first term of Eq. \eqref{eq:non-Karmers_SH} is ignored, leaving the latter two terms. This form is sufficient when working within a single $LSJ$ state, as the largest contributions to the splittings come from interactions with crystal field levels within the same multiplet. Summation over the electron configurations of a single multiplet allows for many of the interactions to be related to the total angular momentum operator $\mathbf{J}$, simplifying the form of the spin Hamiltonian; $\boldsymbol{\Lambda}$ can be used to describe several effects that have the same symmetry despite originating from different physical interactions. 

Clearly, for the $J=0$ multiplets of Eu\tplus the $J$-mixing between multiplets is crucial \citeit{Macfarlane1987}. This prevents the electron configuration summation being used to relate the hyperfine interactions to $\mathbf{J}$, hence preventing the construction of $\boldsymbol{\Lambda}$. Instead, we now sum over all possible electronic states $\ket{\Psi} = \ket{\tau L S J M_J}$ of the $4f^N$ configuration for which the matrix elements of $\mathbf{L}+g_s\mathbf{S}$ and $\mathbf{N}$ are non-zero.

The first correction is the pseudo-nuclear Zeeman. In this case we replace $\boldsymbol{\Lambda}$ by
\begin{equation}
\alpha_{ij} = \sum^{}_{\Psi^\prime}{\frac{\mel**{\Psi}{L_i + g_sS_i}{\Psi^\prime}\mel**{\Psi^\prime}{N_j}{\Psi}}{E_{\Psi^\prime} - E_\Psi}}\,,\label{eq:psuedo_Zeeman}
\end{equation}
where we consider the mixing of levels by the electronic Zeeman and magnetic dipole interactions. Accounting for the pseudo-nuclear Zeeman effect, the nuclear $g$-values can be calculated as
\begin{align}
g_i ={}& g_n\mu_N - 2a_lg_J\mu_B\alpha_{ii}
\end{align}
The second correction is the pseudo-quadrupole. Here we replace $\mathbf{\Lambda}$ by 
\begin{equation}
P_{ij} = \sum^{}_{\Psi^\prime}{\frac{\mel**{\Psi}{N_i}{\Psi^\prime}\mel**{\Psi^\prime}{N_j}{\Psi}}{E_{\Psi^\prime} - E_\Psi}}\,,
\end{equation}
where we consider the second order contribution from the magnetic dipole interaction. The $D$ and $E$ associated with the pseudo-quadrupole are then given by
\begin{align}
D_{pq} ={}& a_l^2\left(\frac{P_{xx}+P_{yy}}{2}-P_{zz}\right)\,,\\
E_{pq} ={}& a_l^2\frac{P_{xx}-P_{yy}}{2}\,.
\end{align}
We also separate the true quadrupole term $\mathbf{T}_Q$ into its two contributions, the lattice $\mathbf{T}_{lat}$ and $4f$ quadrupole $\mathbf{T}_{4f}$, with forms equivalent to Eq. \eqref{eq:sh_Q}. These then have $D$ and $E$ parameters related to crystal field parameters with
\begin{align}
D_{lat} ={}& \frac{3N^2_0}{I(2I-1)}\left(\frac{I(2I-1)}{(I+1)(2I+1)(2I+3)}\right)^{\frac{1}{2}}\,,\\
E_{lat} ={}& \sqrt{\frac{2}{3}}\frac{N^2_2}{N^2_0}D_{lat}\,,
\end{align}
for the lattice contribution, and
\begin{align}
D_{4f} ={}& \frac{3E_Q}{I(2I-1)}\mel**{\psi}{\sum_i\left({{C}_{ei}}\right)^{(2)}_{0}}{\psi}\,,\\
E_{4f} ={}& \sqrt{\frac{2}{3}}\frac{\mel**{\psi}{\sum_i\left({{C}_{ei}}\right)^{(2)}_{2}}{\psi}}{\mel**{\psi}{\sum_i\left({{C}_{ei}}\right)^{(2)}_{0}}{\psi}}D_{4f} \,,
\end{align}
for the $4f$ contribution. Here, $\ket{\psi}$ is the wave function for the state in question, accounting for crystal field mixing.

The spin Hamiltonian for any $J$-mixing dependent system is then given by
\begin{equation}
\begin{aligned}
{ H} ={}& -\mathbf{B}\cdot\left(g_n\mu_N\mathbf{1} 
       -2a_lg_J\mu_B\boldsymbol{\alpha}\right)\cdot\mathbf{I}\\
       &-\mathbf{I}\cdot\left(a_l^2\mathbf{P} + \mathbf{T}_{lat} 
       +\mathbf{T}_{4f}\right)\cdot\mathbf{I}\,. \label{eq:jmix_sh}
\end{aligned}
\end{equation}
Due to the $J$-mixing dependence of the interactions, the principal axes of the quadrupole terms of Eq. \eqref{eq:jmix_sh} are not necessarily aligned. Instead, each contribution takes the form of Eq. \eqref{eq:sh_Q} with a unique set of $D$, $E$, and Euler angles. The contributions are summed in a common frame to form a total quadrupole tensor, with principal axes given by $D_{tot}$ and $E_{tot}$, which are not necessarily aligned with the principal axes of any of the individual quadrupole terms. This is in contrast to the nuclear Zeeman tensor which has principal axes that are aligned with the pseudo-nuclear Zeeman contribution as the pure nuclear Zeeman is isotropic.

\section{Method}\label{sec:method}
Crystal field calculations were performed using M. F. Reid's F-shell empirical program suite in conjunction with S. Horvath's \textit{pycf} program for low symmetry crystal field parameter fitting \citeit{Horvath2016}. A truncated set of matrix elements was used to reduce computation time following the approach of Carnell \etal \citeit{Carnall1976} by first diagonalizing the free-ion Hamiltonian (Eq. \ref{eq:free_ion}) using estimates of the free-ion parameters \citeit{Carnall1989}, and then using a truncated set of eigenvectors of this diagonalization to generate free-ion and crystal field matrix elements in the intermediate coupled basis. We chose to truncate the matrix elements at \num{30} free-ion multiplets out of the total $295$, resulting in $272$ crystal field levels out of the total $3003$, with energies up to $\SI{\sim33000}{\per\centi\metre}$.

For each of the Eu$^{3+}$ centers analyzed, first the free-ion parameters $E_\text{avg}$, $F^k$, and $\zeta_{4f}$ along with the crystal field parameters $B^{k}_q$ required for the site symmetry were fitted to available crystal field energy levels (see Table I in Supplementary Materials). This was done to ensure a satisfactory fit to the crystal field levels before including the hyperfine interactions as adding the nuclear spin means a sixfold increase in the number of states and in the computation time. Next, the crystal field parameters were refitted using the previous parameters as initial values at the same time as fitting the hyperfine parameters $a_l$, $E_Q$, and $N^2_q$. Whilst the lattice quadrupole parameters $N^2_q$ have a fixed relationship to the crystal field parameters $B^2_q$, theoretically requiring the addition of only a single scaling parameter to account for shielding effects, we instead used free parameter for each $q$ term. This was done to avoid propagating error from the crystal field parameters onto the hyperfine structure, which is discussed in Section \ref{sec:discussion}. Instead, bounds were placed on $N^2_q$ such that the appropriate sign and approximate relative magnitudes were correctly imposed by the $B^2_q$ parameters.

For this second round of fitting, the fitting data was the crystal field energy levels, and the hyperfine splittings of the \eug state and the \eue state (when available) calculated for $100$ magnetic field directions defined on a spiral:
\begin{equation}
\mathbf{B} = \begin{bmatrix}
             B_0\sqrt{1-t^2}\cos(6\pi t)\\
             B_0\sqrt{1-t^2}\sin(6\pi t)\\
             B_0 t
             \end{bmatrix}\label{eq:spiral}
\end{equation}
where $B_0 =\SI{400}{\milli\tesla}$. We fitted to hyperfine splittings and not absolute frequencies so that any error in the energy of the crystal field levels was not propagated into the hyperfine errors. This way, the hyperfine splittings could still be fitted even if there was a deviation of the crystal field levels away from the experimental values.

In crystal field theory, there always exist sets of crystal field parameters that produce equivalent fits to energy level structure for a given site, but different wave functions and ordering of $M_J$ levels within the multiplets. These sets do not produce the same Zeeman-hyperfine splittings, so this data is needed to constrain the crystal field values. To avoid the fit getting stuck in a local minimum within the crystal field parameter set two methods were employed; a basin-hopping algorithm \citeit{Wales1997,Wales1999} was used to allow the fit to jump between equivalent sets, and the weighting of the hyperfine and crystal field levels was chosen such that both contributed approximately equally to calculated error. 

The fit was accomplished using the multi-Hamiltonian method \citeit{Horvath2016} where a crystal field Hamiltonian is generated for each magnetic field step then diagonalized concurrently. In this method, the parameters can be fitted using experimental data for both the zero-field crystal field levels and Zeeman-hyperfine splittings simultaneously.

\section{Results}
We studied three different $^{151}$Eu\tplus sites of different symmetry for which sufficient crystal field and hyperfine data exists in the literature. These were the C$_{4v}$ and C$_{3v}$ sites in CaF$_2$, and the C$_2$ site in \eucl. Table \ref{tab:cf_params} lists the fitted crystal field values for each site.
\begin{table}[htb]
\caption{\label{tab:cf_params}Fitted crystal field parameters $(\SI{}{\per\centi\metre})$ for the three Eu\tplus sites studied here. Where a parameter is blank, that parameter does not occur for the site symmetry.}
\begin{tabular*}{\columnwidth}{l@{\extracolsep{\fill}}ccc}\hline\hline
Parameter & \caf:C$_{4v}$ & \caf:C$_{3v}$ & \multicolumn{1}{c}{\eucl} \Tstrut\\ \hline
E$_{avg}$ & $64154$   & $64232$     & $63887$ \Tstrut\\ 
$F_2$   & $83479$     & $83317$     & $82962$ \\
$F_4$   & $60024$     & $59405$     & $59580$ \\
$F_6$   & $42506$     & $42658$     & $42801$ \\
$\zeta$ & $1333$      & $1337$      & $1333$
\vspace{0.15cm} \\
$B^2_0$ & $680$       & $2122$      & $70$\\
$B^2_2$ & \mathemdash & \mathemdash & $241 + 196i$\\
$B^4_0$ & $-852$      & $1697$      & $-334$ \\
$B^4_2$ & \mathemdash & \mathemdash & $335+465i$ \\
$B^4_3$ & \mathemdash & $-2095$     & \mathemdash \\        
$B^4_4$ & $-994$      & \mathemdash & $-316-175i$ \\
$B^6_0$ & $1317$      & $-657$      & $700$ \\
$B^6_2$ & \mathemdash & \mathemdash & $308-396i$ \\
$B^6_3$ & \mathemdash & $-913$      & \mathemdash \\
$B^6_4$ & $-1311$     & \mathemdash & $727-432i$ \\
$B^6_6$ & \mathemdash & $1366$      & $-269+175i$
\vspace{0.15cm} \\
$a_l$   & $0.0405$    & $0.0433$    & $0.0386$ \\
${E}_Q$   & $-0.0490$   & $-0.0497$   & $-0.0408$\\
$N^2_0$ & $-0.00619$  & $-0.0111$   & $-0.000853$\\   
$N^2_2$ & \mathemdash & \mathemdash & $-0.00466-0.00225i$ \Bstrut\\\hline\hline
\end{tabular*}
\end{table}

\subsection{\texorpdfstring{Eu$^{3+}$:CaF$_\mathbf{2}$ C$_\mathbf{4\boldsymbol{v}}$ Center}{Eu3+:CaF2 C4v Center}}
\begin{figure}
\centering
\includegraphics[width=\columnwidth]{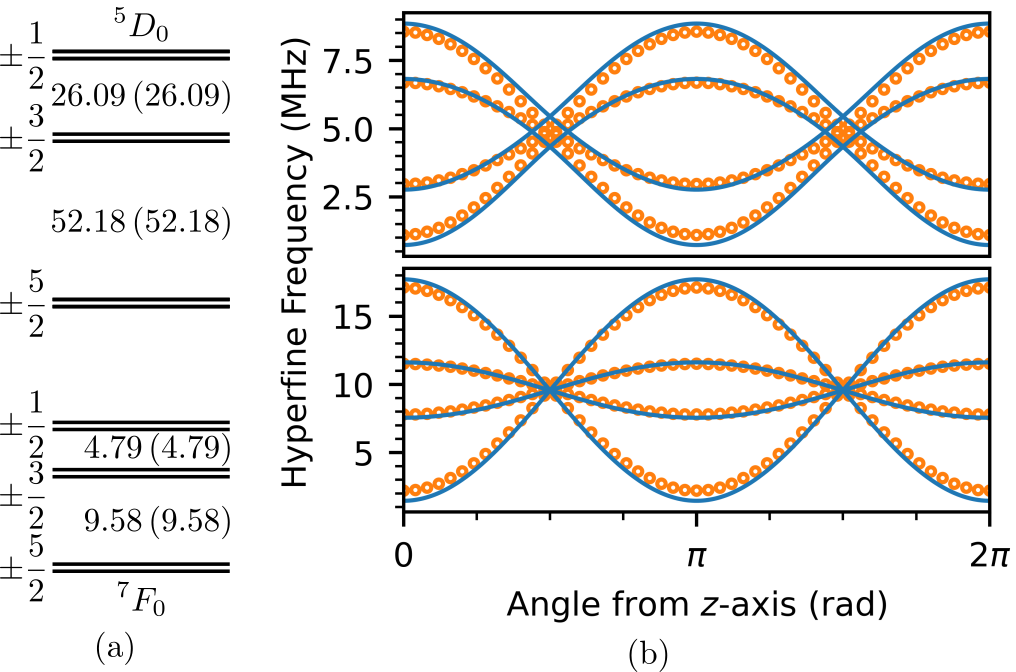}
\caption{\label{fig:c4v}(a) Calculated zero field splittings of the Eu\tplus:\caf C$_{4v}$ site with experimental splittings given in brackets. (b) Calculated (blue) and experimental (orange) Zeeman splittings of the \eug $\pm\frac{3}{2}\rightarrow\pm\frac{1}{2}$ (top) and $\pm\frac{5}{2}\rightarrow\pm\frac{3}{2}$ (bottom) hyperfine transitions for a magnetic field of $\SI{400}{\milli\tesla}$ rotated in the $xz$ plane. Experimental points are calculated from published experimental spin Hamiltonian parameters \citeit{Silversmith1992}.}
\end{figure}

In the C$_{4v}$ site of \caf, Eu\tplus substitutes for Ca$^{2+}$, with charge compensation arising from an interstitial neighboring F$^-$ ion. There are values in the literature for the Zeeman-hyperfine splittings of the \eug and \eue states \citeit{Silversmith1992}, the energy levels \citeit{Wells2001}, and the fitted crystal field parameters \citeit{Wells2001}. Here, we re-fitted the crystal field parameters in the full model using the published experimental energy levels \citeit{Wells2001}. Fitted crystal field parameters are given in Table \ref{tab:cf_params}, and experimental and calculated spin Hamiltonian parameters given in Table \ref{tab:c4v_sh_params}. The re-fitted crystal field parameters are within $\sim10\%$ of the values given in \citeit{Wells2001}. This small change is expected as our fit gives equal weight to the Zeeman-hyperfine splittings, which are most strongly influenced by the positions of the crystal field levels in the \state{7}{F}{1} and \state{7}{F}{2} multiplets. Figure \ref{fig:c4v} shows the comparison between experimental and fitted hyperfine structure for the \eug and \eue states. The Zeeman structure is displayed for fields in a circle in the $xz$ plane:
\begin{equation}
\mathbf{B} = \begin{bmatrix}
             B_0\cos(2\pi t)\\
             B_0\sin(2\pi t)
             \end{bmatrix}\,.\label{eq:zcircle}
\end{equation}
\begin{table}[htb]
\caption{\label{tab:c4v_sh_params} Spin Hamiltonian parameters for the $^{151}$Eu\tplus:\caf C$_{4v}$ site calculated from crystal field fitting compared to experimental spin Hamiltonian parameters \citeit{Silversmith1992}. In that work, signs of the $g$-values could not be determined, so we list them with `$\pm$' here.}
\begin{tabular*}{\columnwidth}{l@{\extracolsep{\fill}}Y{-2.3}Y{-2.3}Y{-2.5}Y{-2.2}}\hline\hline
& \multicolumn{2}{c}{\eug} & \multicolumn{2}{c}{\eue} \Tstrut\\ \cmidrule(lr){2-3} \cmidrule(lr){4-5}
& \multicolumn{1}{c}{Calc} & \multicolumn{1}{c}{Expt} & \multicolumn{1}{c}{Calc} & \multicolumn{1}{c}{Expt} \\ \hline
$D_{pq}$(\si{\mega\hertz})        & 0.198   & \mathemdash & 0.00265 & \mathemdash \Tstrut\\
$D_{4f}$(\si{\mega\hertz})        & 10.968  & \mathemdash & 0.661   & \mathemdash \\
$D_{lat}$(\si{\mega\hertz})       & -13.562 & \mathemdash & -13.708 & \mathemdash \\
$D_{tot}$(\si{\mega\hertz})       & -2.395  & -2.415      & -13.045 & -13.05 \\
$g_x$(\si{\mega\hertz\per\tesla}) & -0.936  & \pm0.212    & 9.646   & \mathemdash \\
$g_y$(\si{\mega\hertz\per\tesla}) & -0.936  & \pm0.212    & 9.646   & \mathemdash \\
$g_z$(\si{\mega\hertz\per\tesla}) & 5.075   & \pm4.657    & 9.761   & \mathemdash\Bstrut\\\hline\hline
\end{tabular*}
\end{table}

The C$_{4v}$ site of \caf is a good illustration of the established qualitative understanding of hyperfine structure in the $J=0$ levels of Eu\tplus, which we briefly summarize. The zero field hyperfine structure in these levels arises from the combination of the lattice quadrupole interaction and effects due to the admixture of $J=2$ crystal field levels into the \eug and \eue wave functions by the crystal field \citeit{Elliott1957, Shelby1981}. This mixing is only large in the ground state as the separation of the $J=2$ levels (the \state{7}{F}{2} multiplet) is much smaller than in the excited state (the \state{5}{D}{2} multiplet). This is clear from the fitted wave functions of the \eug and \eue states (ignoring contributions $<1$\%):
\begin{align}
\Psi(\eug) ={}& -0.99\ket{\eug,0}+0.12\ket{\state{7}{F}{2},0}\,,\\
\Psi(\eue) ={}& +1.00\ket{\eue,0}\,.
\end{align}
This mixing gives rise to the $4f$ contribution to the spin Hamiltonian quadrupole term, which is only significant in the \eug state. However, since this term is of opposite sign to the lattice quadrupole term, the result is smaller quadrupole tensor in the \eug state. 

For this C$_{4}$ site, Fig. \ref{fig:c4v}(a) shows that the agreement between experimental and calculated zero field splittings is excellent. The quantitative parameter results (Table \ref{tab:c4v_sh_params}) also agree well with the qualitative description above. In the \eue state, the lattice contribution to the quadrupole term is over 95\%, whereas in the ground state, roughly equal magnitude, opposite sign $4f$ and lattice contributions result in a quadrupole $<20$\% of the excited state. In both states, the pseudo-quadrupole contribution to the quadrupole tensor is near negligible. This term arises from interactions with $\Delta M_J = 0,1$ levels. For $J\neq0$ states, the pseudo-quadrupole can be significant since these interactions exist within the multiplet. However, for the $J=0$ levels, the large separations to the interacting \state{7}{F}{1} and \state{5}{D}{1} states largely suppress this term. 

While the zero field structure of the $J=0$ levels is determined by mixing with $J=2$ multiplets, the Zeeman structure in a magnetic field is understood to be determined by the small amount of mixing with the $J=1$ multiplets. This mixing allows for non-zero matrix elements for the electronic Zeeman interaction between the nuclear states \citeit{Bleaney1973}. While in $J\neq 0$ states this pseudo-nuclear Zeeman term typically entirely dominates the bare moment, the large spacing to the $J=1$ multiplets means the contribution is much smaller for $J=0$ states \citeit{Elliott1957}. Again, the contribution is typically only significant in the \eug state due to the smaller separation to \state{7}{F}{1}, and again, it is of similar size and opposite sign to that of the true nuclear Zeeman effect. This means $g$-values are typically smaller in the \eug state compared to the \eue state. Depending on the proximity of the \state{7}{F}{1} levels, the pseudo-nuclear Zeeman contribution can be large enough that the sign of the $g$-values is reversed relative to the bare nuclear magnetic moment.

We see from Fig. \ref{fig:c4v}(b) and Table \ref{tab:c4v_sh_params} that, as expected, in \eue the predicted $g$-values are only slightly reduced from the bare moment of \SI{10.58}{\MHz\per\tesla} by the small amount of mixing with \state{5}{D}{1}. In the ground state, the $g$-values are substantially reduced by the mixing. The predicted values in this level do differ from the experimental numbers. This discrepancy can be attributed to the $\sim5$~\% deviation of the calculated crystal field splittings of the \state{7}{F}{1} from the experimental values, since the pseudo-nuclear term is highly sensitive to the separation of the \state{7}{F}{1} and \eug levels. Nevertheless, the agreement is sufficient to resolve the sign ambiguity in the experimental $g$ values: $g_{x,y}$ is negative, and $g_z$ is positive.

\subsection{\texorpdfstring{Eu$^{3+}$:CaF$_2$:O$^{2-}$ C$_\mathbf{3\boldsymbol{v}}$ Center}{Eu3+:CaF2:O2- C3v Center}}
\begin{figure}
\centering
\includegraphics[width=\columnwidth]{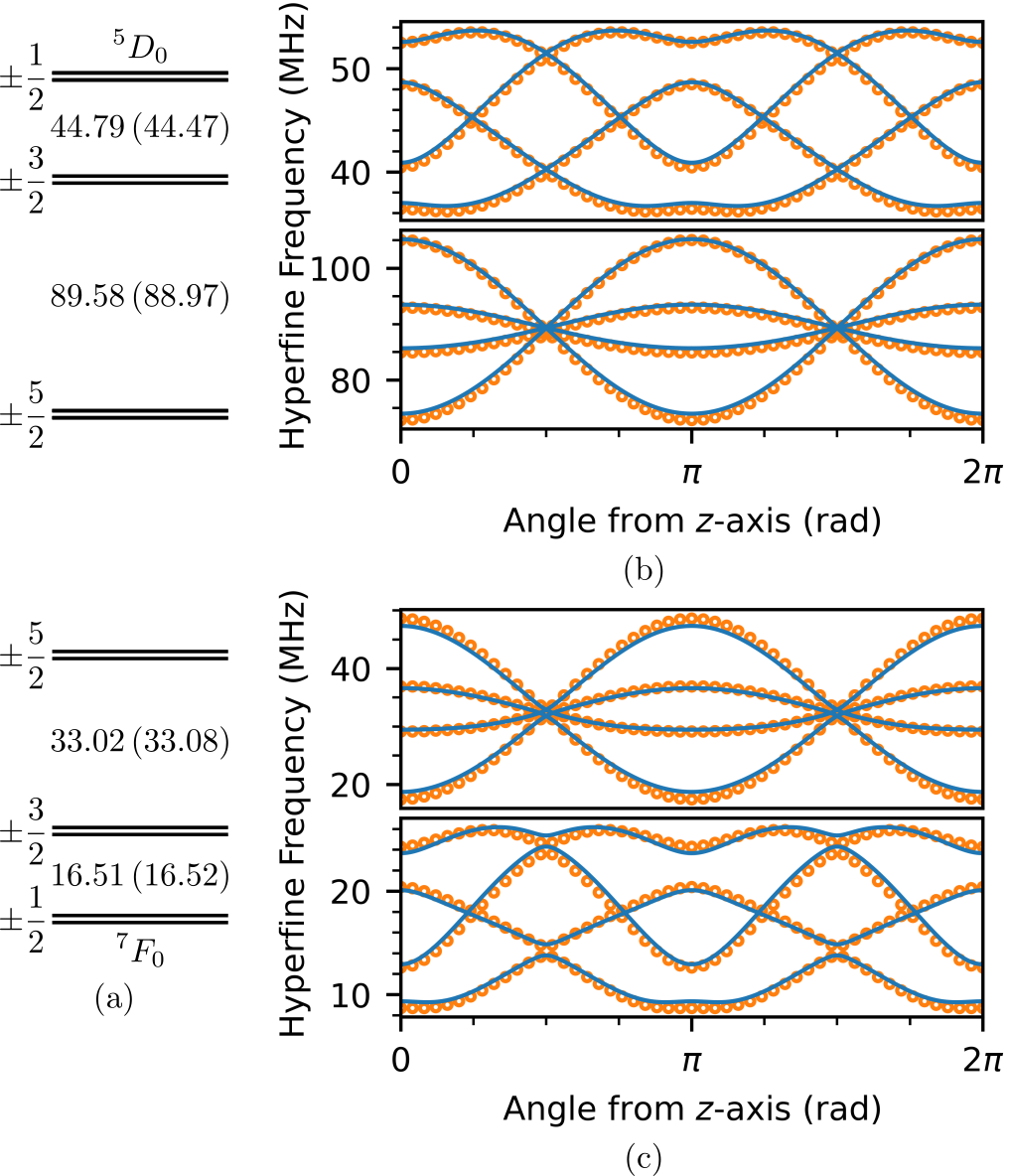}
\caption{\label{fig:c3v}(a) Calculated zero field splittings of the Eu\tplus:\caf C$_{3v}$ site with experimental splittings given in brackets. (b) Calculated (blue) and experimental (orange) Zeeman splittings of the \eue $\pm\frac{3}{2}\rightarrow\pm\frac{1}{2}$ (top) and $\pm\frac{5}{2}\rightarrow\pm\frac{3}{2}$ (bottom) hyperfine transitions for a magnetic field of $\SI{400}{\milli\tesla}$ rotated in the $x-z$ plane. (c) Calculated (blue) and experimental (orange) Zeeman splittings of the \eug $\pm\frac{3}{2}\rightarrow\pm\frac{5}{2}$ (top) and $\pm\frac{1}{2}\rightarrow\pm\frac{3}{2}$ (bottom) hyperfine transitions for a magnetic field of $\SI{400}{\milli\tesla}$ rotated in the $xz$ plane. Experimental points are calculated from published experimental spin Hamiltonian parameters \citeit{Silversmith1986a}.}
\end{figure}

In the C$_{3v}$ site of \caf, denoted the G$1$ center, Eu\tplus substitutes for Ca$^{2+}$ with charge compensation arising from a substitutional O$^{2-}$ in a neighboring F$^-$ site. Zeeman-hyperfine splittings for this site are available for both the \eug ground and \eue excited states \citeit{Silversmith1985f, Silversmith1985d, Silversmith1986d, Silversmith1985a, Silversmith1986a, Radlinski1986}, but despite these extensive hyperfine studies, the observed crystal field levels are restricted to few \state{7}{F}{J} and \state{5}{D}{J} multiplets \citeit{Silversmith1985f}. Fitting a set of crystal field parameters to the available energy levels is further complicated by the very strong crystal field caused by the nearby oxygen atom \citeit{Silversmith1985d} which makes assignment of the \state{7}{F}{1} and \state{7}{F}{2} levels difficult \citeit{Silversmith1985d, Silversmith1985f}. As there has been no prior crystal field fits to the G$1$ center and the ordering of the \state{7}{F}{1,2} levels is unknown, we estimated a set of initial crystal field parameters by hand using the \state{5}{D}{1} and \state{5}{D}{2} splittings. This set was then refined by fitting to all of the available crystal field and Zeeman-hyperfine data for the site. The ordering of the \state{7}{F}{1} and \state{7}{F}{2} levels could then be inferred from the crystal field calculation, demonstrating that the unusually large crystal field causes an overlap of the \state{7}{F}{1} and \state{7}{F}{2} multiplets. Fitted crystal field parameters are given in Table \ref{tab:cf_params}, and experimental and calculated spin Hamiltonian parameters given in Table \ref{tab:c3v_sh_params}. Figure \ref{fig:c3v} demonstrates the very good agreement between fitted and experimental splittings, with Zeeman structure calculated using Eq. \eqref{eq:zcircle}.
\begin{table}[h!tb]
\caption{\label{tab:c3v_sh_params}Spin Hamiltonian parameters for the $^{151}$Eu\tplus:\caf C$_{3v}$ site calculated from crystal field fitting compared to experimental spin Hamiltonian parameters \citeit{Silversmith1986a}.}
\begin{tabular*}{\columnwidth}{l@{\extracolsep{\fill}}Y{-2.3}Y{-1.3}Y{-2.5}Y{-2.3}}\hline\hline
& \multicolumn{2}{c}{\eug} & \multicolumn{2}{c}{\eue} \Tstrut\\ \cmidrule(lr){2-3} \cmidrule(lr){4-5}
& \multicolumn{1}{c}{Calc} & \multicolumn{1}{c}{Expt} & \multicolumn{1}{c}{Calc} & \multicolumn{1}{c}{Expt} \\ \hline
$D_{pq}$(\SI{}{\mega\hertz})        & 0.610     & \mathemdash & 0.00809 & \mathemdash \Tstrut\\
$D_{4f}$(\SI{}{\mega\hertz})        & 31.988    & \mathemdash & 2.170   & \mathemdash \\
$D_{lat}$(\SI{}{\mega\hertz})       & -24.344   & \mathemdash & -24.573 & \mathemdash \\
$D_{tot}$(\SI{}{\mega\hertz})       & 8.256     & 8.26        & -22.395 & -22.23 \\
$g_x$(\SI{}{\mega\hertz\per\tesla}) & -9.658    & -8.998      & 9.504   & 9.527 \\
$g_y$(\SI{}{\mega\hertz\per\tesla}) & -9.658    & -8.998      & 9.504   & 9.527 \\
$g_z$(\SI{}{\mega\hertz\per\tesla}) & 8.948     & 9.739       & 9.751   & 10.056 \Bstrut\\ \hline\hline
\end{tabular*}
\end{table}

Unlike the C$_{4v}$ center, the C$_{3v}$ center does not conform well to the normal picture for $J=0$ energy level structure because the unusually large crystal field causes considerable mixing between multiplets. In the \eue state the quadrupole interaction $D = \SI{-22.395}{\mega\hertz}$ is still primarily due to the lattice quadrupole $D_{lat} = \SI{-24.573}{\mega\hertz}$, but the contribution from the $4f$ quadrupole $D_{4f} = \SI{2.170}{\mega\hertz}$ is double that for the C$_{4v}$ center. This is due to the large $B^2_0 = \SI{2122}{\per\centi\metre}$ term mixing the \eue with the \state{5}{D}{2} state despite these multiplets being separated by $\approx\SI{4000}{\per\centi\metre}$. The calculated wave functions
\begin{align}
\Psi(\eug)={}&+0.92\ket{\eug,0}-0.35\ket{\state{7}{F}{2},0}+0.11\ket{\state{7}{F}{4},0} \nonumber\\
&+0.10\ket{\state{7}{F}{4},-3}-0.10\ket{\state{7}{F}{4},3}\,,\\
\Psi(\eue)={}&-1.00\ket{\eue,0}+0.04\ket{\state{5}{D}{2},0}\,,
\end{align}
demonstrate the considerable mixing compared to the C$_{4v}$ site.

In the \state{7}{F}{} multiplet, the crystal field splittings of the \state{7}{F}{1} and the \state{7}{F}{2} states generated by the $B^2_0$ and $B^4_q$ terms are so large that, as described above, the multiplets are no longer separated and the ordering of states is not obvious. In particular, a doublet state of \state{7}{F}{2} at $\SI{758}{\per\centi\metre}$ is lower in energy compared to the singlet state of \state{7}{F}{1} at $\SI{865}{\per\centi\metre}$ (see Supplementary Material Table I). The low lying \state{7}{F}{2} mixes strongly with \eug resulting in a large $4f$ quadrupole contribution $D_{4f} = \SI{31.988}{\mega\hertz}$. This is larger than the lattice quadrupole contribution $D_{lat} = \SI{-24.344}{\mega\hertz}$, resulting in a total quadrupole interaction of $D_{tot} = \SI{8.256}{\mega\hertz}$. This reversal of the sign of the quadrupole interaction results in a reversal of the ordering of the nuclear spin states compared to the \eue state. 

The $g$-values of the C$_{3v}$ site do follow a similar pattern to the C$_{4v}$ case. In the excited state, the $g$-values are only slightly reduced from the bare moment with good agreement between calculation and experiment. Despite the strong crystal field, the splittings between the \eue and \state{5}{D}{1} states are still large. The influence of the large crystal field is seen more strongly in the ground state. The large splitting within the \state{7}{F}{1} state pushes the $M_J = \pm 1$ doublet far closer to the \eug state. The resulting large mixing via the magnetic dipole interaction results in a pseudo-nuclear Zeeman effect almost twice the size of the bare moment in the $x$ and $y$ directions. Conversely, since the \state{7}{F}{1} singlet state is pushed up, the $z$ component of the pseudo-nuclear Zeeman effect is small ($<\SI{2}{\MHz\per\tesla}$) as it arises from mixing with this term. The result is $g_{x,y}$ and $g_z$ values of similar magnitude but opposite sign, agreeing well with the experimental values. Differences in the calculated and experimental \eug $g$-values are due to inaccuracies in the fitted crystal field levels as the mixing of states is proportional to the splitting as evidenced in Eq. \eqref{eq:psuedo_Zeeman}.

Previous works have attempted to explain the large pseudo-nuclear Zeeman contribution observed in \eug of the C$_{3v}$ center \citeit{Silversmith1985f, Silversmith1986a}. This was done without the use a crystal field calculation and it was thought that $J$-mixing of \state{7}{F}{2} into \eug would not exceed $1\%$. Based on this assumption it was concluded that $J$-mixing could not explain the observed deviations of the electronic Zeeman and magnetic dipole matrix elements from their free-ion values. However, by performing a complete crystal field calculation we have been able to show that $J$-mixing is far larger than previously thought, close to $13\%$, which is sufficient to explain the observed pseudo-nuclear Zeeman effect.

\subsection{\texorpdfstring{EuCl$_\mathbf{3}$.6H$_\mathbf{2}$O C$_\mathbf{2}$ Center}{EuCl3 C2 Center}}
\eucl is a stoichiometric crystal in which the Eu\tplus ion occupies a single site of C$_2$ symmetry. Several papers over many years have studied the electronic and Zeeman-hyperfine structure \citeit{Longdell2006, Binnemans1997, Stump1999, Stump1994, Hellwege1951c, Kahle1959, Martin1998}, and it has more recently been investigated for quantum information applications \citeit{Ahlefeldt2013, Ahlefeldt2013b, Ahlefeldt2020}.

\begin{figure}
\centering
\includegraphics[width=\columnwidth]{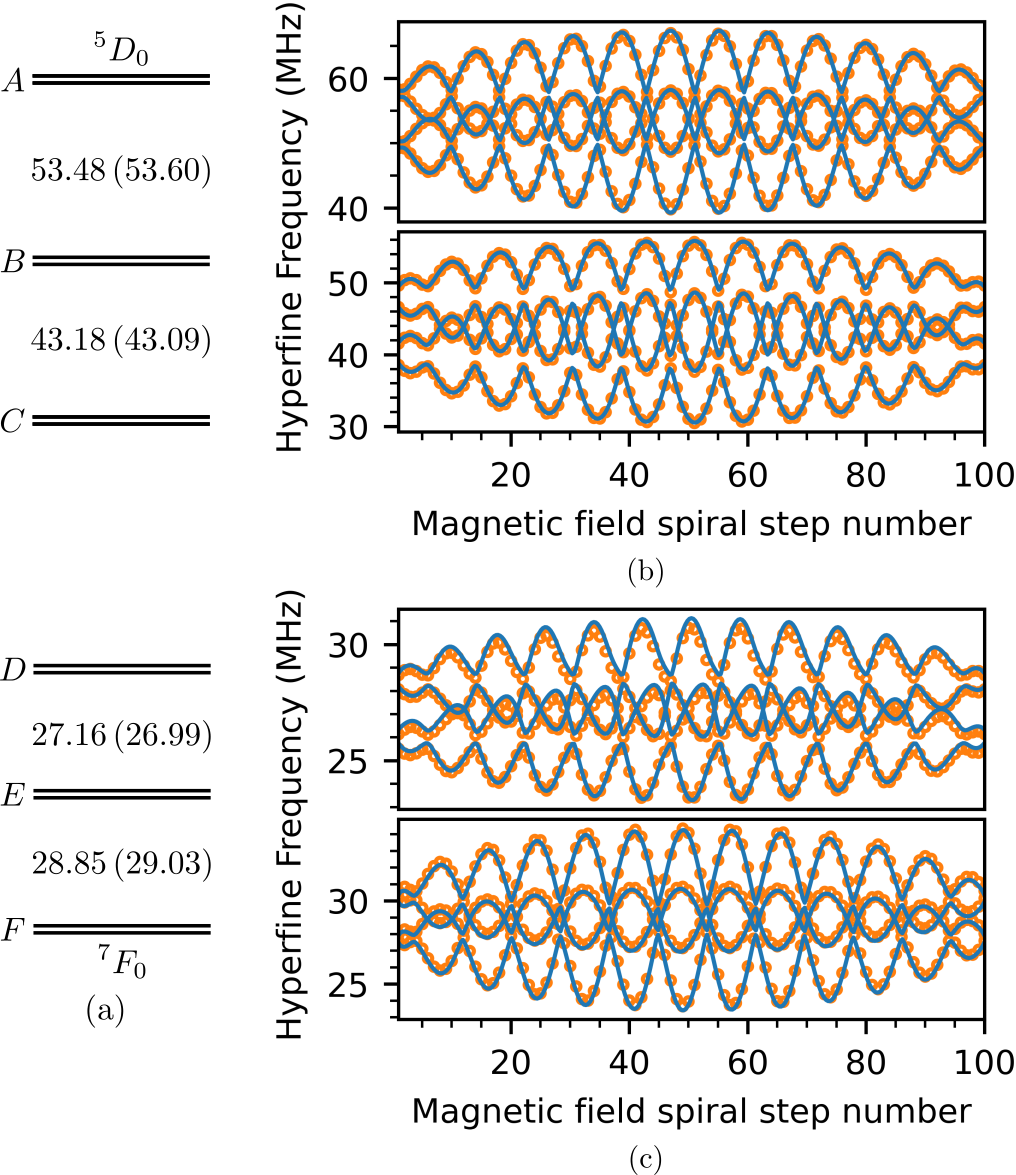}
\caption{\label{fig:eucl_rotation_patterns}(a) Calculated zero field splittings of the \eucl C$_2$ site with experimental splittings given in brackets. (b) Calculated (blue) and experimental (orange) Zeeman splittings of the \eue $B\rightarrow A$ (top) and $C\rightarrow B$ (bottom) hyperfine transitions for a magnetic field spiral of $\SI{400}{\milli\tesla}$. (c) Calculated (blue) and experimental (orange) Zeeman splittings of the \eug $E\rightarrow D$ (top) and $F\rightarrow E$ (bottom) hyperfine transitions for a magnetic field spiral of $\SI{400}{\milli\tesla}$. Experimental points are calculated from published experimental spin Hamiltonian parameters \citeit{Ahlefeldt2013}.}
\end{figure}

The non-axial site symmetry means that additional crystal field terms must be included: 15 crystal field parameters compared to the 5 (6) for the C$_{4v}$ (C$_{3v}$) sites. The large number of crystal field parameters caused previous workers to fit the \eucl crystal field in the higher C$_{2v}$ symmetry \citeit{Binnemans1997}, but it is precisely the additional crystal field terms that generate the correctly shaped spin Hamiltonian quadrupole and Zeeman tensors. The quadrupole and Zeeman tensors arising from each of the interactions described in Section \ref{sec:spinhamil} are no longer axial, and the principal axes of these tensors need only coincide along the C$_2$ direction. Here, we use a selection of the energy levels from Ref. \citeit{Binnemans1997} along with some new \state{7}{F}{J} levels determined from fluorescence measurements to fit the site (see Supplementary Materials Table I) using C$_2$ symmetry and the method described in Section \ref{sec:method}. Fitted crystal field parameters are given in Table \ref{tab:cf_params}, and experimental and calculated spin Hamiltonian parameters given in Table \ref{tab:eucl_sh_params}. The fitted crystal field parameters are notably different to the those of Ref. \citeit{Binnemans1997}. This is due to a combination of the lowering of symmetry from C$_{2v}$ to C$_2$ and the use of different fixed free-ion parameters. Despite the low symmetry, the agreement between experimental and calculated hyperfine splittings is excellent, as shown in Fig. \ref{fig:eucl_rotation_patterns}. To display the Zeeman splitting in this non-axial site, we have calculated the structure for the magnetic field spiral in Eq. \eqref{eq:spiral}.
\begin{table}[h!tb]
\caption{\label{tab:eucl_sh_params}$^{151}$\eucl spin Hamiltonian parameters calculated from crystal field fitting compared to experimental spin Hamiltonian parameters \citeit{Ahlefeldt2013}. Note that we have transformed the spin Hamiltonian parameters of Ref. \citeit{Ahlefeldt2013} into the standard electron paramagnetic formalism: the $zyz$ Euler rotation convention, Eq. \eqref{eq:euler_rotation}. Further, we have chosen the opposite set of equivalent spin Hamiltonian $E$, $g_x$ and $g_y$ parameters to match the crystal field fit as described in the text.}
\begin{tabular*}{\columnwidth}{l@{\extracolsep{\fill}}Y{-2.5}Y{-2.5}Y{-2.6}Y{-2.5}}\hline\hline
& \multicolumn{2}{c}{\eug} & \multicolumn{2}{c}{\eue} \Tstrut\\ \cmidrule(lr){2-3} \cmidrule(lr){4-5}
& \multicolumn{1}{c}{Calc} & \multicolumn{1}{c}{Expt} & \multicolumn{1}{c}{Calc} & \multicolumn{1}{c}{Expt} \\ \hline
{$D_{pq}$(\SI{}{\mega\hertz})}        & 0.0550  & \mathemdash  & 0.000239  & \mathemdash  \Tstrut\\
{$E_{pq}$(\SI{}{\mega\hertz})}        & 0.0938  & \mathemdash  & 0.000725  & \mathemdash  \\
{$\gamma_{pq}$(\SIUnitSymbolDegree)}  & -68.33  & \mathemdash  & -68.33    & \mathemdash  \\
{$D_{4f}$(\SI{}{\mega\hertz})}        & 2.109   & \mathemdash  & 0.0497    & \mathemdash  \\
{$E_{4f}$(\SI{}{\mega\hertz})}        & 4.275   & \mathemdash  & 0.199     & \mathemdash  \\
{$\gamma_{4f}$(\SIUnitSymbolDegree)}  & -68.01  & \mathemdash  & -68.01    & \mathemdash  \\
{$D_{lat}$(\SI{}{\mega\hertz})}       & -1.866  & \mathemdash  & -1.875    & \mathemdash  \\
{$E_{lat}$(\SI{}{\mega\hertz})}       & 9.263   & \mathemdash  & 9.284     & \mathemdash  \\
{$\gamma_{lat}$(\SIUnitSymbolDegree)} & 12.89   & \mathemdash  & 12.89     & \mathemdash  \\
{$D_{tot}$(\SI{}{\mega\hertz})}       & 0.29718 & 0.35692     & -1.82529  & -1.85868 \\
{$E_{tot}$(\SI{}{\mega\hertz})}       & 5.29024 & 5.29026     & 9.08941   & 9.08595 \\
{$\gamma_{Qtot}$(\SIUnitSymbolDegree)}& 5.43    & 2.93        & 12.76     & 13.22 \\
{$g_x$(\SI{}{\mega\hertz\per\tesla})} & 4.115   & \pm4.029    & 9.690     & 9.666 \\
{$g_y$(\SI{}{\mega\hertz\per\tesla})} & -1.723  & \mp1.504    & 9.647     & 9.293 \\
{$g_z$(\SI{}{\mega\hertz\per\tesla})} & 2.857   & 3.006       & 9.686     & 10.025 \\
{$\gamma_M$ (\SIUnitSymbolDegree)}    & 21.81   & 22.69       & 19.30     & 3.1\Bstrut\\\hline\hline
\end{tabular*}
\end{table}

Again, the excited state quadrupole is dominated by the lattice term, where the additional parameter $E$ accounts for the asymmetry of this non-axial site. The calculated wave functions
\begin{align}
\Psi(\eug)={}&(-0.11-0.99i)\ket{\eug,0}+(0.05+0.04i)\ket{\state{7}{F}{2},2} \nonumber\\
&+(-0.04+0.05i)\ket{\state{7}{F}{2},-2}\,, \\
\Psi(\eue)={}&(-0.34-0.94i)\ket{\eue,0}\,,
\end{align}
show the strong mixing of the $M_J=\pm2$ \state{7}{F}{2} levels into \eug due to the large $B^2_2$ crystal field parameter. Similarly, there is no mixing of the $M_J=0$ \state{7}{F}{2} level into \eug by the much weaker $B^2_0$ crystal field parameter. As the rank two crystal field parameters are directly related to $D$ and $E$, this explains the large asymmetry of the spin Hamiltonian quadrupole contributions. There is close agreement between the experimental and calculated values for both the principal axes values and orientation of the tensor. The ground state has comparable $4f$ and lattice quadrupole contributions. These two terms have different orientations of the principle axes, leading to the overall quadrupole tensor in the ground state being misaligned from the excited state by \SI{10}{\degree}, in reproducing the rotation seen in the experimental data. The fit also resolved the ambiguity in the sign of $E$: as described in section \ref{sec:spinhamil}, in a C$_2$ site the sign of $E$ is poorly defined in a spin Hamiltonian model because a reversal of sign can be compensated for by rotating the $M$ tensor to redefine the orientation of the $x$ and $y$ axes. However, the sign of $E$ has physical significance in the crystal field model, and the fit shows that, indeed, the sign of $E$ is negative and the experimental spin Hamiltonian parameters need to be transformed $E\rightarrow -E, (g_x,g_y)\rightarrow (g_y,g_x)$, $\gamma\rightarrow\gamma\pm\ang{90}$.

The excited state $g$-values are again, only slightly reduced from the bare moment. The larger, \SI{10}{\degree}, discrepancy with the experimental orientation of the Zeeman tensor is unsurprising: given the tensor is nearly isotropic, its orientation is not well defined in either the experimental data or the theoretical calculation. In the ground state, the agreement between experimental data and calculation is close, with the non-axial symmetry giving rise to three unique $g$-values. From the point of view of the crystal field, this occurs because the additional crystal field parameters split the $M_J=\pm 1$ doublet. The matrix elements of $L_x+2S_x$ and $N_x$ are large between \eug and the $M_J=+1$ singlet of \state{7}{F}{1}, and the matrix elements of $L_y+2S_y$ and $N_y$ are large between \eug and the $M_J=-1$ singlet of \state{7}{F}{1}. As the $M_J=-1$ singlet is the lowest lying \state{7}{F}{1} state with a gap of $\SI{301}{\per\centi\metre}$, the pseudo-nuclear contribution is largest for fields applied in the $y$ direction, resulting in a large negative contribution to the $g$-value, sufficient to reverse the sign of the term. Similarly, the smaller $x$ contribution from the $M_J=+1$ singlet $\SI{428}{\per\centi\metre}$ away results in a $g_x$ value closer to the bare moment, with $g_z$ taking a value between these two extremes. Previous work \citeit{Ahlefeldt2013} was only able to determine that $g_x$ and $g_y$ had opposite signs; this fit resolves this sign ambiguity.

The Zeeman-hyperfine splittings were also calculated for $^{153}$Eu\tplus by scaling the the electric quadrupole moment $Q$ and the nuclear magnetic moment $g_n$ by the \iso{153}Eu/\iso{151}Eu ratio, $2.56$ and $0.43$ respectively \citeit{Stone2005}. This is not sufficient to match the experimental hyperfine structure, since there is an isotope shift in the crystal field levels which slightly $\mathcal{O}(1\%)$ alters the electronic contributions to the hyperfine levels. This shift is $\sim\SI{220}{\mega\hertz}$ on the $\eug\rightarrow\eue$ transition but can be expected to be much larger on $J\neq 0$ transitions. To account for these differences, we fine-tuned the full set of crystal field parameters to obtain the fitted values in Table \ref{tab:eucl_sh_params_153}.
\begin{table}[h!tb]
\caption{\label{tab:eucl_sh_params_153}$^{153}$\eucl spin Hamiltonian parameters calculated from crystal field fitting compared to experimental spin Hamiltonian parameters \citeit{Ahlefeldt2013}. Note that we have transformed the spin Hamiltonian parameters of Ref. \citeit{Ahlefeldt2013} into the standard electron paramagnetic formalism: the $zyz$ Euler rotation convention, Eq. \eqref{eq:euler_rotation}. Further, we have chosen the opposite set of equivalent spin Hamiltonian $E$, $g_x$ and $g_y$ parameters to match the crystal field fit as described in the text.}
\begin{tabular*}{\columnwidth}{l@{\extracolsep{\fill}}Y{-2.6}Y{-2.6}Y{-2.6}Y{-2.3}}\hline\hline
& \multicolumn{2}{c}{\eug} & \multicolumn{2}{c}{\eue}\Tstrut\\ \cmidrule(lr){2-3} \cmidrule(lr){4-5}
& {Calc} & {Expt} & {Calc} & {Expt} \\ \hline
$D_{tot}$(\SI{}{\mega\hertz}) & 0.813097 & 0.791815 & -4.815367 & -4.80 \Tstrut\\
$E_{tot}$(\SI{}{\mega\hertz}) & 13.648917 & 13.660932 & 23.209892 & 23.21 \\
$\gamma_{Qtot}$(\SIUnitSymbolDegree) & 3.21 & 2.93 & 12.19 & 13.22 \\
$g_x$(\SI{}{\mega\hertz\per\tesla}) & 1.878 & \pm1.794 & 4.315 & 4.269 \\
$g_y$(\SI{}{\mega\hertz\per\tesla}) & -0.422 & \mp0.673 & 4.298 & 4.105 \\
$g_z$(\SI{}{\mega\hertz\per\tesla}) & 1.381 & 1.359 & 4.315 & 4.428 \\
$\gamma_M$ (\SIUnitSymbolDegree) & 23.08 & 22.69 & 20.47 & 3.1 \Bstrut\\ \hline\hline
\end{tabular*}
\end{table}

\begin{figure}
\centering
\includegraphics[width=\columnwidth]{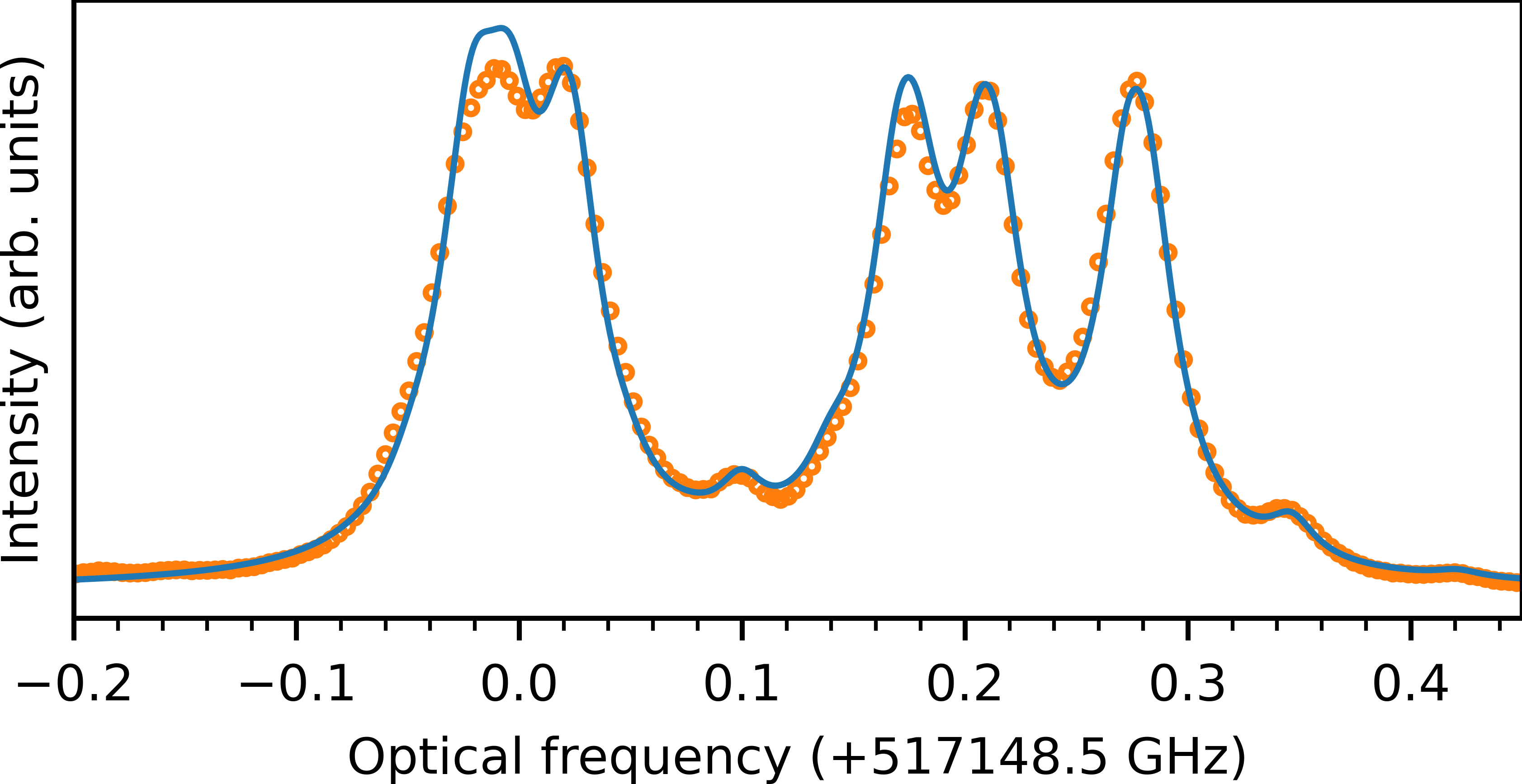}
\caption{\label{fig:oscillator_strengths}Simulated and experimental spectra of the $\eug\rightarrow\eue$ nuclear state transitions of \iso{151}Eu\tplus and \iso{153}Eu\tplus in \eucl. Simulated spectra was calculated using the function $\Gamma(f) = A_0\alpha(f)e^{-\alpha(f)A_1}$ which accounts for absorption of the light in this optically thick sample before reaching the point in the crystal from which emission was collected. We used a Lorentzian line shape $\alpha$ with FWHM line width \SI{0.0122}{\giga\hertz}, and an \iso{153}Eu\tplus transition offset of $\SI{0.220}{\giga\hertz}$.}
\end{figure}

Relative oscillator strengths between \eug and \eue may be calculated using Eq. \eqref{eq:oscillator_strengths} (Table \ref{tab:oscillator_strengths}), or the equivalent expression for the full crystal-field Hamiltonian. Since both electronic states are singlets, the relative intensities are determined by the overlap of the nuclear spin states between the ground and excited states. This calculation provides an independent test of the model with data that was not used in the fitting process. Since experimental oscillator strengths are not directly available, we instead compared the excitation spectrum these values generate for the \eutrans transition with the experimental spectrum for Eu\iso{35}Cl$_3$.6H$_2$O in Fig. \ref{fig:oscillator_strengths}. The simulated spectrum has only three free parameters: the overall amplitude, the isotope shift (which is too small to be accurately reproduced by the crystal field model), and a parameter accounting for absorption. Given this, the very good agreement confirms the calculated oscillator strengths are accurate.
\begin{table}[h!tb]
\caption{\label{tab:oscillator_strengths} Oscillator strengths of the \eucl $\eug\rightarrow\eue$ nuclear state transitions. State labels follow those of Fig. \ref{fig:eucl_rotation_patterns}. Calculated values are given by the overlap of the crystal field nuclear wave functions and experimental values are taken from \citeit{Ahlefeldt2013}.}
\begin{tabular*}{\columnwidth}{l@{\extracolsep{\fill}}Y{1.4}Y{1.4}Y{1.4}Y{1.4}}\hline\hline
& \multicolumn{2}{c}{$^{151}$Eu\tplus} & \multicolumn{2}{c}{$^{153}$Eu\tplus}\Tstrut\\ \cmidrule(lr){2-3} \cmidrule(lr){4-5}
& {Calc} & {Expt} & {Calc} & {Expt} \\ \hline
$F\rightarrow C$ & 0.9608 & 0.9300 & 0.9446 & 0.9302 \Tstrut\\
$F\rightarrow B$ & 0.0341 & 0.0608 & 0.0483 & 0.0609 \\
$F\rightarrow A$ & 0.0050 & 0.0091 & 0.0071 & 0.0089 \\
$E\rightarrow C$ & 0.0378 & 0.0687 & 0.0541 & 0.0686 \\
$E\rightarrow B$ & 0.9322 & 0.8766 & 0.9031 & 0.8771 \\
$E\rightarrow A$ & 0.0299 & 0.0547 & 0.0428 & 0.0544 \\
$D\rightarrow C$ & 0.0013 & 0.0013 & 0.0013 & 0.0013 \\
$D\rightarrow B$ & 0.0336 & 0.0626 & 0.0486 & 0.0620 \\
$D\rightarrow A$ & 0.9650 & 0.9362 & 0.9501 & 0.9367\Bstrut\\ \hline\hline
\end{tabular*}
\end{table}

\section{Discussion}\label{sec:discussion}
We have shown crystal field theory can be used to accurately calculate Zeeman-hyperfine splittings of the $J=0$ levels in several different Eu\tplus as long as the nuclear Zeeman and lattice electric quadrupole terms are included in the calculation. In particular, the lattice contribution to the nuclear quadrupole can be accurately calculated as shown by the \eue splittings in each material, including calculating the rotation about the symmetry axis (in low symmetry) for both the lattice and $4f$ contributions. The crystal field mixing of the \state{7}{F}{2} and \state{7}{F}{1} levels into \eug is also well reproduced, allowing accurate determination of the hyperfine structure of this state. In the C$_{3v}$ case we showed that the large deviation of the magnetic dipole and electronic Zeeman matrix elements from their free-ion values \citeit{Silversmith1986a} is indeed due to an extraordinarily large $J$-mixing of the \state{7}{F}{2} and \state{7}{F}{4} levels into \eug, resolving an open question in the literature. 


The model fit the hyperfine structure of each material remarkably well despite, in some cases, how few crystal field levels were available to constrain the fits. In fact, this is the result of a unusual property of the low lying levels in Eu\tplus: certain crystal field levels have reduced matrix elements \rme{\Psi}{U^{(k)}}{\Psi} which are large for particular $k$-values and small or non-zero for all other $k$, allowing those levels to be used to constrain the corresponding crystal field parameter. In Eu\tplus, \state{7}{F}{1} is sensitive to $B^{2}_q$, \state{7}{F}{2} is sensitive to $B^{4}_q$, and \state{7}{F}{3}/\state{7}{F}{4} are sensitive to $B^{6}_q$, along with \state{5}{L}{6} to a lesser extent. Whilst there are similar relationships between parameters and individual multiplets and crystal field parameters in other ions (see Ref. \citeit{Gorller-Walrand1996a} for a full list), Eu\tplus is particular as these special levels are low-lying, so are readily observed in fluorescence spectra of the \eue level. Further, their assignment is rarely ambiguous due to the true singlet nature of \eug and \eue levels, and the large separations \state{7}{F}{J} and \state{5}{D}{J} levels. It is only in particularly unusual cases such as the \caf C$_{3v}$ site that \state{7}{F}{} multiplets overlap. For these reasons, Eu\tplus might be a good dopant choice when trying to work out the crystal field fitting parameters of a new host material: since parameters vary only slowly across the rare earth series \citeit{Carnall1989} parameters from Eu\tplus could be used as an initial guess for other rare earth dopants for which assigning observed energy levels is more difficult.

We have emphasized, and demonstrated, that crystal field fits can resolve spin Hamiltonian ambiguities. However, this requires considerable care to be taken during the fitting process, because there are ambiguities in the crystal field fit itself when fitting to only the energies of crystal field levels. There are two types of ambiguity that manifest in crystal field fitting \citeit{Rudowicz1985, Rudowicz2000}. The first are rotations of the crystal field parameters about the symmetry axis, which occur in centers with imaginary components. The rotational ambiguity arises when the relationship between the crystal field and the experimental crystallographic axes is unknown. It is only by constraining the crystal field parameters through measurements that depend on the true orientation of the crystal, such as Zeeman splittings, that we can resolve this ambiguity. The second ambiguity is complete re-orientations of the symmetry axis which can yield as many as six numerically equivalent crystal field parameter sets \citeit{Rudowicz2000}. This is typically only an issue at orthorhombic and lower symmetries \citeit{Rudowicz1985, Rudowicz1986} with multiple non-zero crystal field parameters of every rank. Whilst the the calculated splittings are equivalent for these different sets, the calculated wave functions will differ. Therefore, it is possible to determine the correct set of crystal field parameters by comparing wave function dependent quantities, such as oscillator strengths and Zeeman splittings, to experimental values.

In Eu\tplus, wave function-dependent information that can be obtained with relative ease. This includes the ordering of the \state{7}{F}{1} $M_J$ levels, and the ordering of the \eue $M_I$ hyperfine states at zero-field. In high symmetry where the only non-zero rank two crystal field parameter is $B^2_0$, the \eue hyperfine state ordering is sufficient due to the relationship between the splittings and the lattice quadrupole (Eq. \eqref{eq:lat_cf_param}). However, in lower symmetry the introduction of additional $B^2_q$ parameters means there is insufficient information from just this ordering,  and knowledge of the ordering the of the \state{7}{F}{1} levels is needed as well. This can be achieved by measuring the polarized absorption or fluorescence of the $\state{7}{F}{1}\longleftrightarrow\state{5}{D}{0}$ transition, or by measuring the \eug nuclear Zeeman $g$-values. Once the ordering of the \state{7}{F}{1} levels is known, these can be used to determine the sign and magnitudes of $B^2_q$ as there is a straightforward relationship between the ordering of the states and the parameters \citeit{Gorller-Walrand1996a}. Restricting the rank two parameters to a single set then restricts the other crystal field parameters. Solving the crystal field parameters sets ambiguities simultaneously solves the spin Hamiltonian sign ambiguities.

 
Comparison of the experimental and calculated crystal field highlights a well-known limitation of the standard, one-electron crystal field model used here: the model omits two-electron perturbation correlation effects (referred to as the correlation crystal field) \cite{Gorller-Walrand1996a}. These additional effects are known to be important for correctly predicting orderings and splittings of the \state{5}{D}{J} states of Eu\tplus \citeit{Moune1989}, and, indeed, we see larger deviations between experimental and calculated parameters for those levels (Supplementary Materials Table I). Additionally, omitting these effects can cause a change in the free-ion parameters $F_k$, artificially pushing the \state{7}{F}{J} states down in energy. This is the likely source of the discrepancies of the calculated \eug and \state{7}{F}{1} levels in \eucl, with each shifted downward by significantly more than the standard deviation $\sigma=\SI{15}{\per\centi\meter}$. It is not normally practical to include the correlation crystal field because doing so requires a large number of additional parameters, up to $637$ for the lowest site symmetry \citeit{Garcia1995}. Therefore, effects that would otherwise be accounted for by the correlation crystal field are instead absorbed by the single-electron crystal field. This results in crystal field parameters that are not necessarily a true representation of the crystal field potential.

We made one modification to the model to minimize the effect of this small distortion of the crystal field parameters. Eq. \eqref{eq:lat_cf_param} shows that the lattice electric quadrupole $N^2_q$ parameters should have a fixed dependence on the single-electron crystal field $B^2_q$ parameters. Instead, we allowed the lattice quadrupole parameters $N^2_q$ to vary semi-independently of $B^k_q$, which avoids propagating the distortion of the crystal field parameters through to the hyperfine structure. In this way, the $N^2_q$ parameters are more representative of the true crystal field than the $B^2_q$ parameters. This ensures a satisfactory fit to both the hyperfine and crystal field structure, whilst maintaining consistency between the crystal field and quadrupole parameters.



\section{Conclusions}
We have calculated Zeeman-hyperfine splittings of Eu\tplus in three crystallographic centers, the C$_{4v}$ and C$_{3v}$ sites of \caf as well as the C$_2$ site of \eucl, using a complete crystal field model, with excellent agreement between experimental and calculated values. This was achieved by including the generally omitted lattice quadrupole interaction and nuclear Zeeman interaction, which are important for accurate calculations of hyperfine splittings for non-Kramers singlet states. We have also demonstrated accurate crystal field calculations of oscillator strengths in low symmetry (C$_2$) by using all parameters required for the exact symmetry of the site.

\appendix*
\section{}
The $zyz$ Euler rotation convention used is given by
\begin{equation}
\begin{aligned}
R\left(\varphi,\theta,\psi\right) = {}&
\begin{bmatrix}
 \cos\varphi & \sin\varphi & 0\\
-\sin\varphi & \cos\varphi & 0\\
 0           & 0           & 1\\
\end{bmatrix}\times
\begin{bmatrix}
\cos\theta & 0 & -\sin\theta\\
0          & 1 &  0\\
\sin\theta & 0 &  \cos\theta\\
\end{bmatrix}\\
&\times\begin{bmatrix}
 \cos\psi & \sin\psi & 0\\
-\sin\psi & \cos\psi & 0\\
 0        & 0        & 1\\
\end{bmatrix}\, .
\end{aligned}\label{eq:euler_rotation}
\end{equation}

\bibliographystyle{apsrev4-2}
\bibliography{writeups-europium_2020}
\end{document}